\documentclass[onecolumn]{aastex62}

\usepackage{amsmath}

\accepted{May 13, 2022}
\submitjournal{ApJL}

\shorttitle{}
\shortauthors{\c{S}ahin and Antolin}

\graphicspath{{./}{figures/}}
\def\ion#1#2{#1$\;${\small\rm\@{#2}}\relax}

\newcommand{\longacknowledgment}{The authors would like to thank the anonymous referee for the constructive comments and acknowledge T. Schad for his help with the RHT algorithm. P.A. acknowledges funding from his STFC Ernest Rutherford Fellowship (No. ST/R004285/2). This study used openly available IRIS and SDO data at https://www.lmsal.com. IRIS is a NASA small explorer mission developed and operated by LMSAL with mission operations executed at NASA Ames Research Center and major contributions to downlink communications funded by ESA and the Norwegian Space Centre.  SDO is part of NASA’s Living With a Star Program.} 

\begin{document}

\title{Prevalence of Thermal Non-Equilibrium over an Active Region}

\author[0000-0003-3469-236X]{Seray \c{S}ahin}
\author[0000-0003-1529-4681]{Patrick Antolin}
\affiliation{Department of Mathematics, Physics and Electrical Engineering, Northumbria University, Newcastle Upon Tyne, NE1 8ST, UK}

\begin{abstract}
Recent observations have shown that besides the characteristic multi-million degree component the corona also contains a large amount of cool material called coronal rain, whose clumps are $10-100$ times cooler and denser than the surroundings and are often organised in larger events termed showers. Thermal instability (TI) within a coronal loop in a state of thermal non-equilibrium (TNE) is the leading mechanism behind the formation of coronal rain but no investigation on showers exists to date. In this study, we conduct a morphological and thermodynamic multi-wavelength study of coronal rain showers observed in an active region (AR) off-limb with IRIS and SDO, spanning chromospheric to transition region and coronal temperatures. Rain showers were found to be widespread across the AR over the 5.45-hour observing time, with average length, width and duration of 27.37$\pm 11.95$~Mm, 2.14$\pm 0.74$~Mm, and 35.22$\pm20.35$~min, respectively. We  find a good correspondence between showers and the cooling coronal structures consistent with the TNE-TI scenario, thereby properly identifying coronal loops in the `coronal veil', including the  strong expansion at low heights and an almost zero expansion in the corona. This agrees with previous work suggesting that the observed zero expansion in the EUV is due to specific cross-field temperature distribution. We estimate the total number of showers to be $155\pm 40$, leading to a TNE volume of $4.56\pm [3.71]\times10^{28}$ $\text{cm}^3$, i.e. on the same order of the AR volume. This suggests a prevalence of TNE over the AR indicating strongly stratified and high-frequency heating on average.

\end{abstract}


\section{Introduction} \label{sec:intro}

Coronal rain is one of the most mesmerizing features in the solar atmosphere, commonly found in quiescent \citep{schrijver2001,antolin2012} and flaring active regions  \citep{scullion2016,Jing_2016NatSR...624319J}, and also over quiet Sun regions in hybrid prominence/coronal rain complexes \citep{LiL_2018ApJ...864L...4L,ChenH_2022AA...659A.107C}. It is a cooling phenomenon believed to originate due to thermal instability within a coronal loop in a state of thermal non-equilibrium (TNE) \citep[\, and references therein]{antolin2020,Antolin_Froment_2022}. TNE describes a specific behaviour of the plasma within a magnetic flux tube that is unable to reach thermal equilibrium. The plasma undergoes a non-linear cyclic behaviour (limit cycles), where each cycle is characterised by a (short) heating and a (long) cooling phase, also called evaporation and condensation phases, respectively. Numerical simulations show that the most efficient way of achieving a TNE state is through strongly stratified and high-frequency heating \citep[i.e., with heating events repeating on a time scale shorter than the cooling timescale of the structure;][]{Muller_2003AA...411..605M, LiX_2022ApJ...926..216L}. During the heating, chromospheric evaporation occurs, and the loop gets dense and starts to radiate strongly. During the cooling, catastrophic cooling can occur, leading to thermal instability (TI) locally in the corona \citep{Claes:2021wg}. TI is thought to generate cool and dense condensations called rain clumps, which subsequently fall as coronal rain towards the surface under the action of various forces \citep{Oliver_2014ApJ...784...21O, Fang_2015ApJ...807..142F}. Periodic coronal rain is therefore expected and is observed at the end of each TNE cycle \citep{auchere2018,froment2020}. On the other hand, there are cases predicted by numerical simulations of TNE cycles with incomplete condensations or no condensations at all for which the temperature of the plasma does not go down to chromospheric or transition region temperatures \citep{froment2018ApJ}.

Coronal rain shows a clumpy ($\sim$300 km width) and elongated ($\sim$700 km length) \citep{antolin2012} multi-thermal ($<10^3->10^5$~K) structure along the magnetic field lines and densities varying between 2$\times$ 10$^{10}$ and 2.5$\times$ 10$^{11}$ cm$^{-3}$ \citep{antolin2015, froment2020}. A major characteristic of coronal rain is that it often occurs at similar times over a significantly wide structure (relative to the clump width), with a cross-field length scale of a few Mm, thereby defining a `shower’, observed both in observations \citep{antolin2012} and simulations \citep{fang2013}. It has been suggested that the shower’s evolution and morphology reflect a coherent heating function in neighboring field lines, thus defining a coronal volume with a similar thermodynamic evolution that we may associate with the concept of coronal loop. Furthermore, the cross-field quasi-simultaneous occurrence seen in numerical simulations calls for a syncing mechanism acting across the field, which has been explained through the mechanism of  sympathetic cooling. With 2.5D MHD simulations, \citet{fang2013} showed that this mechanism can come from fast mode perturbations. However, there is no observational study to date that properly quantifies  the properties of showers, nor the spatial and temporal scales over which the sympathetic cooling occurs. Quantifying showers is important since it provides a measure of the coronal volume involved in TNE and subject to TI over an active region. 

Because of the optically thin nature of the solar corona, the observational identification of a coronal loop is strongly ambiguous. The loop concept is further ill-defined because of the fuzzy boundaries of what constitutes a coherent structure in terms of its thermodynamic evolution. This is best exemplified in 3D MHD modelling, where continuous magnetic connectivity changes are obtained \citep{gudiksen2005, Malanushenko_2022ApJ...927....1M}. Yet, coherent evolution on a global scale clearly exists, and a clear example of this is the long-period intensity pulsation \citep{Auchere_2014AA...563A...8A, Froment_2015ApJ...807..158F}. Debate also exists on the apparent lack of expansion of loop structures in the EUV \citep{Watko2000, Klimchuk2000SoPh..193...53K,DeForest2007,Fuentes2008}, which may be explained by non-trivial density and coronal temperature cross-field distribution in expanding flux tubes \citep{Peter2012}. 

In this study, we provide a statistical investigation of showers over an active region. We present their morphological properties in Section \ref{sec:morphology} and investigate the cooling behaviour of the plasma to understand the relation with TNE-TI in Section \ref{sec:thermal}. Finally, in Section \ref{sec:TNE}, we estimate the TNE volume in the active region. A discussion and conclusion are given in Section \ref{sec:results}.

\begin{figure*}[ht]
\includegraphics[width=\textwidth]{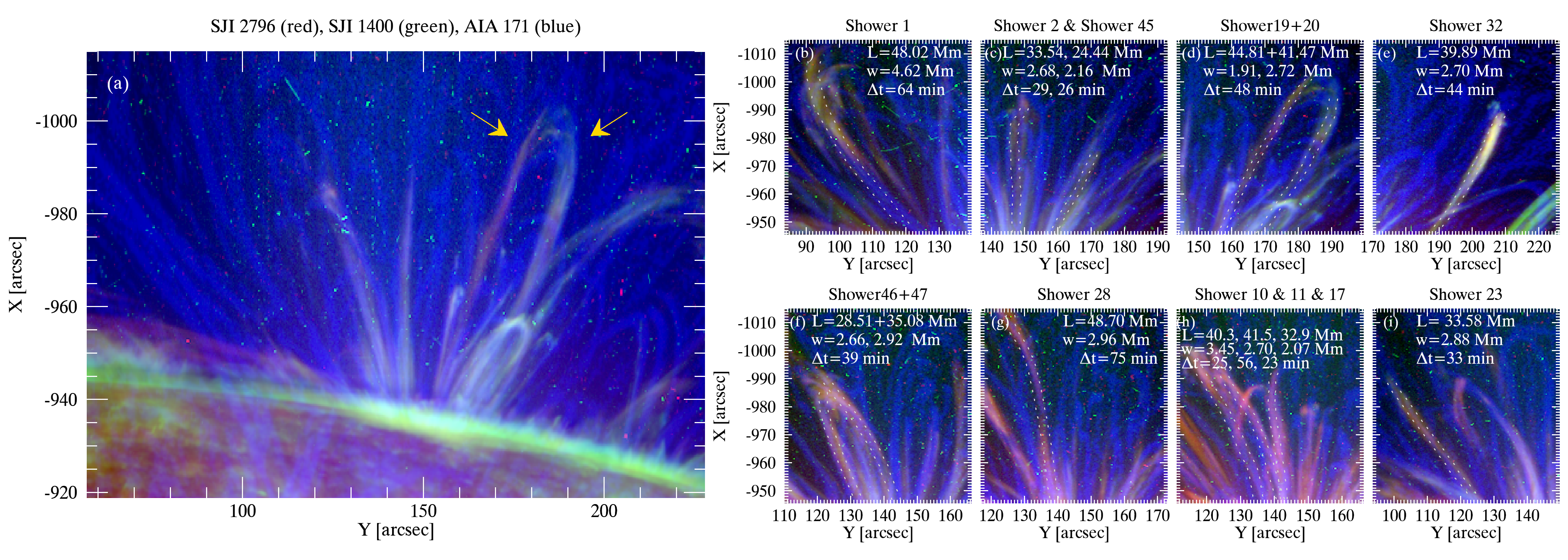}
\caption{a) Composite image of a studied active region at the East limb of the Sun on 2 June 2017,  co-observed by IRIS/SJI~2796 (red, dominated by the \ion{Mg}{II}~k line), IRIS/SJI~1400 (green, dominated by the Si~IV~1402.77~\AA\, line) and SDO/AIA~171 (blue, dominated by the Fe~IX~171.073~\AA\, line). The image corresponds to a variance of 30 images over intervals 08:59:54 UT and 09:16:05 UT for each channel. We applied the Multiscale Gaussian Normalization (MGN) technique \citep{Morgan2014} to AIA 171\AA~ in order to enhance the coronal structures. The yellow arrows show two legs (19 and 20, hereafter Shower 19+20) of a traced shower event, for which we study the thermodynamic evolution in Section~\ref{sec:thermal}. b,c,d,e,f,g,h,i) composite images of some detected shower events, with their corresponding width, length and time duration displayed on the top right of each pannel. The dotted curves correspond to the shower contours based on SJI~2796.\label{shower_panels}}
\end{figure*}

\section{Observations and Methods} \label{sec:data}

The active region chosen for this study, shown in Figure \ref{shower_panels}{a}, is NOAA 12661 and was observed on 2 June 2017 between 07:28:00 UT and 12:54:39 UT at the East limb by the Interface Region Imaging Spectrograph \citep[IRIS;][]{depontieu2014} slit-jaw imager (SJI) in 1400~\AA~ and 2796~\AA. We also study the observations of this region in seven channels (304~\AA, 171~\AA, 193~\AA, 211~\AA, 335~\AA, 94~\AA, and 131~\AA) by the Atmospheric Imaging Assembly \citep[AIA;][]{lemen2012} on board the Solar Dynamics Observatory  \citep[SDO;][]{pesnell2012}. The IRIS/SJI field of view (FOV) for this region is $232\arcsec\times182\arcsec$.
The co-alignment routine between the AIA and the SJI channels was achieved by matching the solar limb and on-disk features at similar times between AIA~1600 and SJI~2832, which form in similar low chromospheric conditions. 

\begin{figure*}[ht]
\includegraphics[width=\textwidth]{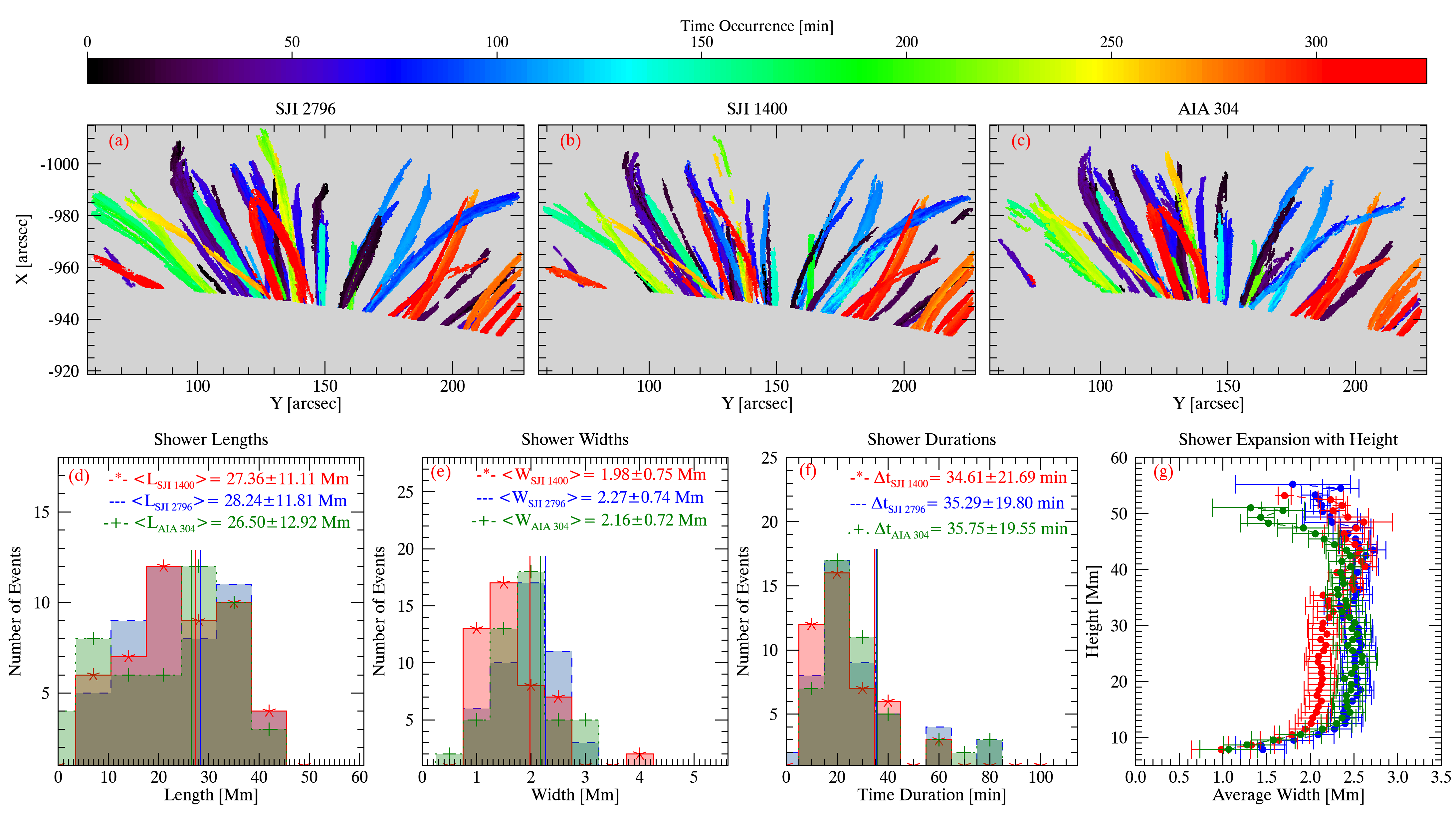}
\caption{Top: Time occurrence of traced shower events in SJI~2796 (a), SJI~1400 (b), and AIA~304 (c). Bottom: 1D Histogram distribution of shower lengths (d), widths (e), and time durations (f), with the corresponding average and standard deviation in the inner caption. Average shower width and standard deviation with height (g). \label{figure1}}
\end{figure*}

We applied an automatic detection routine to trace coronal rain based on the Rolling Hough Transform (RHT) following the procedure of  \citet{schad2017}. The RHT code provides the spatial mean angle, which denotes the spatial inclination in the FOV of the rain's trajectory, and the time occurrence for each rain pixel. A pair of showers may have very similar trajectories but occur at different times. Similarly, they may occur at similar times and overlap each other but have different trajectories. Hence, to identify the shower events we used a semi-automatic algorithm that uses both the spatial mean angle and the time occurrence outputs. This is achieved by creating a new array obtained by the multiplication of the spatial mean angle and the time occurrence data and a shower is defined by a range of values in this trajectory-time domain. This procedure effectively differentiates the cases of spatial or temporal overlap highlighted above. To identify a shower we applied the `region\_grow' routine of the Interactive Data Language (IDL) (common in Medicine science) to this new trajectory-time array. This routine is a space filling algorithm that allows to select a group of pixels within a specific range of values in the domain defined by a standard deviation. The algorithm uses an initial pixel belonging to a shower (as seed) and given a standard deviation (which determines the range of values to look for in the trajectory-time domain) all the pixels belonging to a shower are identified. The standard deviation is carefully selected by visually determining the possible trajectory and time range values. We repeat this process for each shower by selecting 50 starting positions all over the FOV and across the time sequence. The selection is also based on the shower’s visibility in the FOV. Preference was given to significantly isolated events, thus reducing any possible remaining errors due to overlaps. Since most of the rain clumps are visible across all channels, most (if not all) showers can be seen in each channel. 
Considering this, we manually chose 50 shower events. We then analysed their properties in each channel (see panels a, b and c in Figure \ref{figure1}). 

To calculate the shower width, for each pixel within a shower defined by the region\_grow algorithm we draw a perpendicular line to the rain trajectory. The width of the shower at that pixel is then given by the number of adjacent shower pixels.
By repeating this process along the length of the shower we can define the spine of the shower, and piece-wise sum the pixels along the spine to recover the shower length. The shower duration is simply determined by the first and last snapshots in which the shower (or a part of it) is seen in the FOV.

The AIA~304 channel is dominated by \ion{He}{II}~303.8~\AA\, plasma emission at $\log T\approx 5$. However, the channel also includes hotter emission at $\log T\approx 6.17$, mainly from \ion{Si}{XI}~303.32~\AA. The morphology of both emitting structures is strongly different, the former being clumpy since it is emitted by the rain while the latter is diffuse since it is emitted mainly by the surrounding corona. To maximise the rain detection in AIA~304 (and to minimise possible errors) we removed the diffuse emission from the AIA~304 images using the Blind Source Separation technique, following the procedure of \citet{Dudok-de-Wit:2013uw}.

We also use the Differential Emission Measure (DEM) algorithm of  \citet{Cheung2015} based on the Basis Pursuit method to estimate the temperature variation of the loops hosting the showers. Observations taken from six EUV channels of SDO/AIA (94~\AA, 131~\AA, 171~\AA, 193~\AA, 211~\AA, and 335~\AA) were used to calculate the emission measure distribution for each pixel.


\section{Results} \label{sec:results}


\begin{figure*}[ht]
\includegraphics[width=0.9\textwidth]{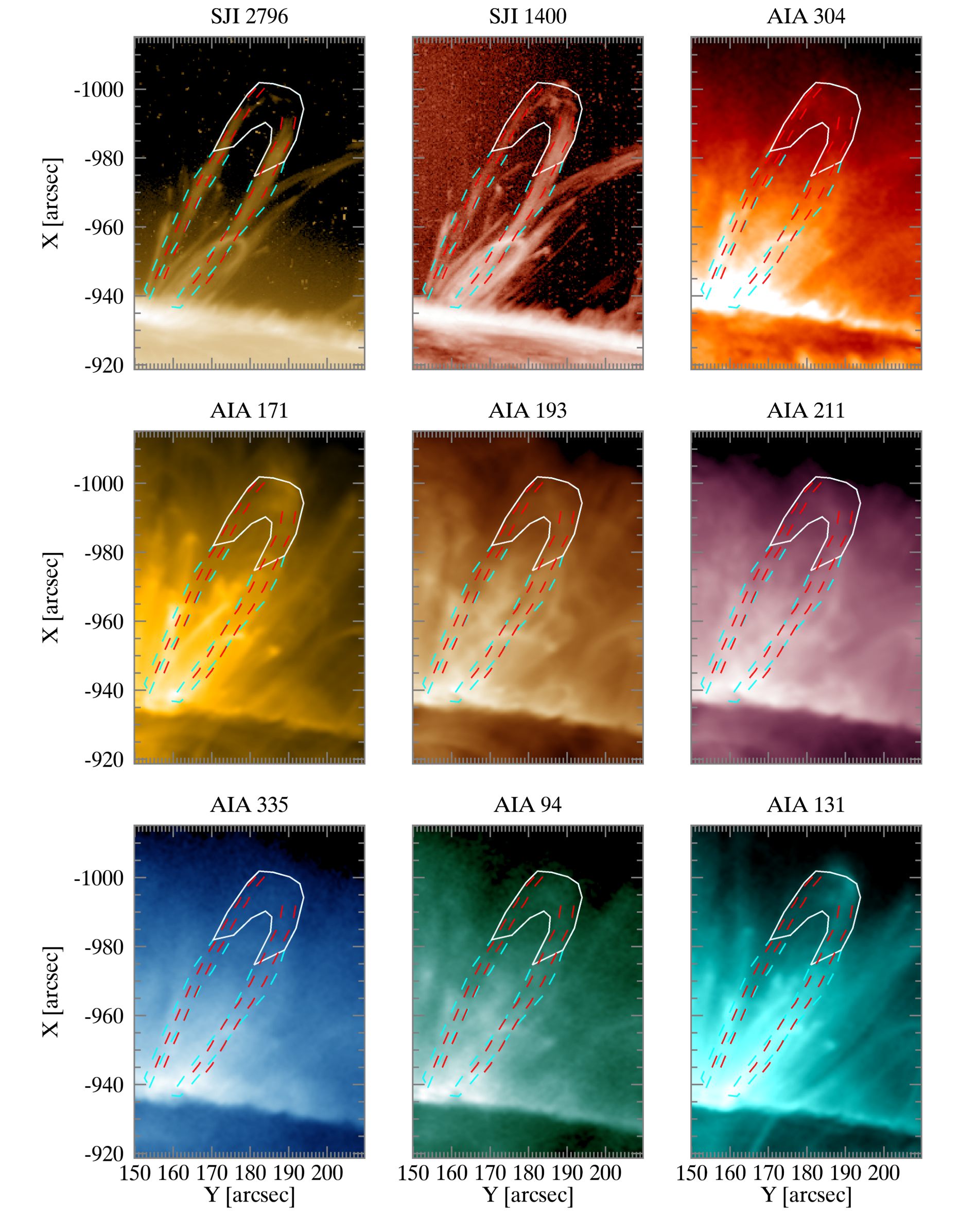}
\caption{SJI and AIA channels of Shower 19+20 (see also panels d and e in Figure \ref{shower_panels}). Each map corresponds to the average over a time duration of 11 minutes. The solid white contour shows the apex of the loop. The cyan contour shows the loop legs derived from 171 (panel a in Figure \ref{shower_panels}), while the red contour corresponds to the shower captured by the region grow algorithm. The accompanying animation shows the SJI (radial filter has been applied to SJIs in order to reduce the intensity of the disc to be able to see more clearly the off-limb structures) and AIA (the 304 images correspond to the original AIA 304, without removal of the diffuse component) channels of Shower$19+20$ over the time span of 111~min.}  \label{sji_aia_channels}
\end{figure*}

\subsection{Morphology of Showers} \label{sec:morphology}

Figure \ref{shower_panels}{a} shows the active region under study combining SJI~2796, SJI~1400, and AIA~171. The panels b to i in the figure show 8 examples of showers with their corresponding length, width and time duration, determined with the methods outlined in section~\ref{sec:data}. These showers have been selected based on their relatively isolated and clean coronal emission (in blue). Note that the dotted lines, based on SJI~2796 contours, match well the coronal structure, which suggests that the loops and showers are very similar in width. 
Furthermore, we note that the showers occur mostly in one of the two loop legs, and only for some the shower encompasses the full loop structure, including the apex. For the few cases in which showers occur along both loop legs we take the average of both in the respective measured quantity (width, length and duration). Based on visual inspection, we estimate showers to occupy a third of the loop length on average.

The top three panels in Figure \ref{figure1} show all the traced 50 shower events in SJI~2796, SJI~1400, and AIA~304, and the colors indicate their time occurrence. Some relatively minor differences are seen between the channels, which are due to differences in intensity (opacity of the line) and contrast with the background, the amount of noise (instrument sensitivity), the spatial resolution of the instrument and also differences in the occurrence with height. 
In any case, the showers can be clearly distinguished and appear ubiquitous over the active region. 
Panels d and e of Figure \ref{figure1} show 1D histogram distributions of the lengths and widths of showers, respectively. We obtain a very similar average shower length and width across the channels. Namely, average lengths of 28.24$\pm 11.81$~Mm for  SJI~2796, 27.36$\pm 11.11$~Mm for SJI~1400, and 26.50$\pm 12.92$~Mm for AIA~304,  and average shower widths of 2.27$\pm 0.74$~Mm, 1.98$\pm 0.75$~Mm, and 2.16$\pm 0.72$~Mm, respectively. Due to the higher noise, lower sensitivity and lower opacity in SJI~1400 and AIA~304 compared to SJI~2796, the lengths and widths appear slightly shorter. A detailed analysis of these differences between the channels is given in \citet{Sahin_rht_2022}.

Showers are relatively long-lived. They have similar time duration distributions across the channels (35.29$\pm 19.80$~Mm for  SJI~2796, 34.61$\pm 21.69$~Mm for SJI~1400, and 35.75$\pm 19.55$~Mm for AIA~304, see Figure \ref{figure1}{f}) and can last as long as 80 minutes. The average width variations with height are shown on panel g in Figure \ref{figure1}, where zero height denotes the solar limb. There is almost no expansion in the upper corona up to 40~Mm height. On the other hand, very strong expansion also be seen below 12~Mm. The rain at the apex of loops is not traced very well by the RHT routine due to the low velocities (particularly when the angle between the loop planes and the line-of-sight is small). This gives rise to strong variations in the measured widths that can be seen above a height of 40~Mm in Figure \ref{figure1}.  


\subsection{Temperature Evolution in Shower} 
\label{sec:thermal}

Many loop structures hosting the showers become bright in the EUV at their apex, prior or during the catastrophic cooling and appearance of coronal rain. An example of this is shown in Figure \ref{sji_aia_channels} and Animation \ref{sji_aia_channels}, where Shower 19$+$20 is displayed (see also panels d and e in Figure \ref{shower_panels}). In order to study the thermodynamic evolution more closely, we focus on shower events that are relatively isolated, thus minimising loop overlapping (Shower 19+20 and Shower 32 in Figure \ref{figure1} on panels d and e, respectively). 

\begin{figure*}[ht]
\includegraphics[width=\textwidth]{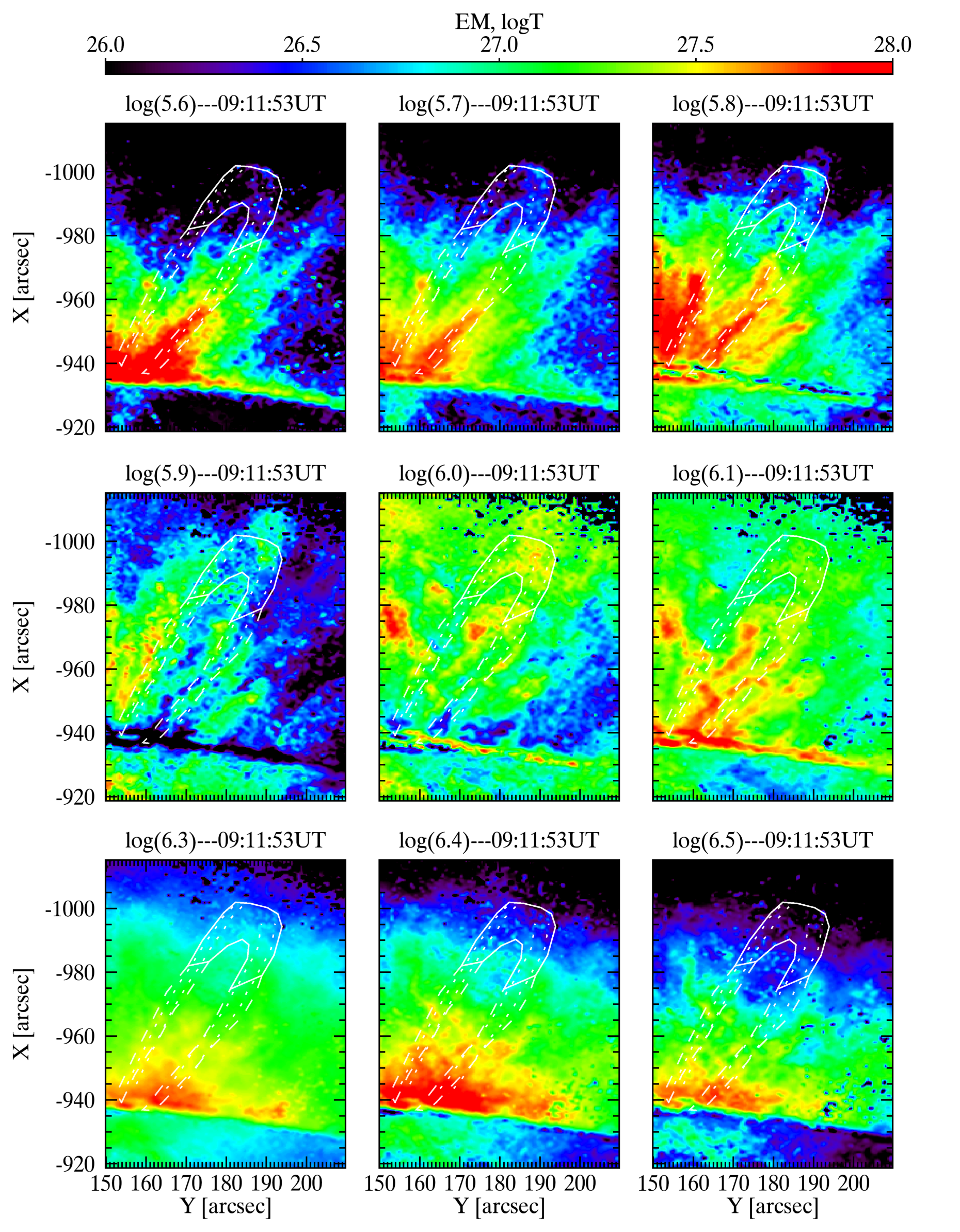}
\caption{EM maps for the same shower in Figure \ref{sji_aia_channels}. The solid white contour shows the apex of the loop. The dashed contour shows the loop legs derived from AIA~171 (panel a in Figure \ref{shower_panels}), while the dotted contour corresponds to the shower captured by the region grow algorithm. The accompanying animation shows the EM maps of Shower$19+20$ for the same duration as in Animation \ref{sji_aia_channels}. \label{dem_maps}}
\end{figure*}

Figure \ref{dem_maps} (and the corresponding animation) presents the DEM results for various temperature bins corresponding to Shower 19+20 (marked in Figure \ref{shower_panels}). The apex of the loop hosting this shower can be identified in almost all EM maps inside the solid-white contour area. The shower event starts at 08:54:29 UT and lasted 48 minutes. However, we investigated the DEM from one hour before the start of the shower. While widespread cooling from coronal temperatures across the FOV can be seen, the cooling is particularly clear in this loop. In the region that is determined by solid white contour (apex of the loop), we have an increase in the DEM in the cooler temperature bins, which constitutes evidence of loop cooling. 

\begin{figure*}[ht]
\includegraphics[width=\textwidth]{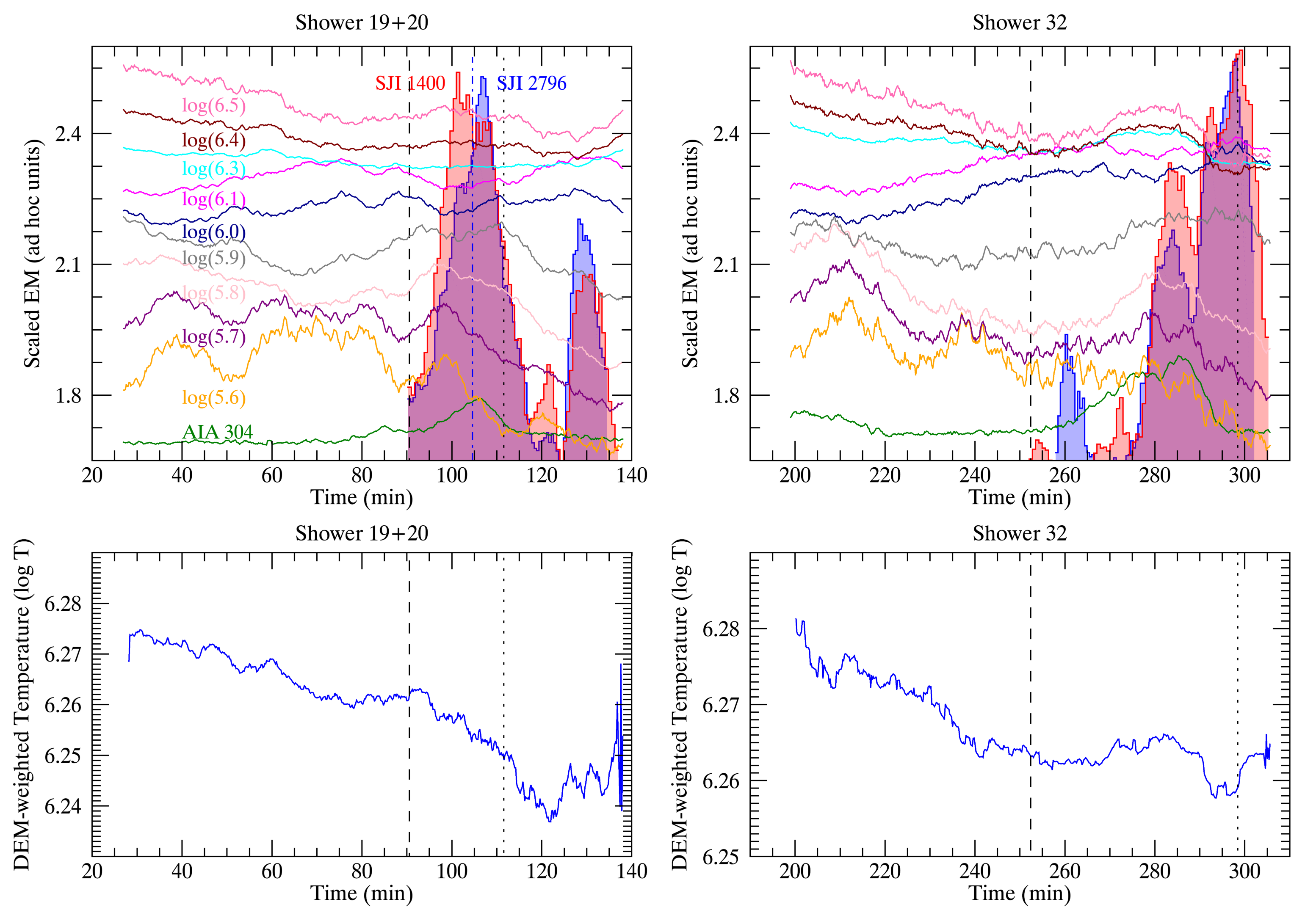}
\caption{Top: Scaled EM plots over the white-solid contour shown in Figures \ref{sji_aia_channels} and \ref{dem_maps} for showers 19+20 (left) and 32 (right). The histograms correspond to the coronal rain pixels (including both legs in the case of 19+20) that are detected with RHT and region grow algorithms in SJI 2796 (blue) and SJI 1400 (red). The legend identifying each curve is indicated in the top left panel (the order and colours are the same for Shower 32). Bottom: DEM-weighted temperature for both showers. The vertical black dashed lines show the start of the shower events. The vertical black dotted line indicates the time when the  shower is not observed anymore at the apex. Blue dashed-dotted vertical line on Shower 19$+$20 plot corresponds to the time of Figures \ref{sji_aia_channels} and  \ref{dem_maps}. \label{em_dem_plots}}
\end{figure*}

To see the cooling trend more clearly, we show in Figure \ref{em_dem_plots} the scaled total DEM over the loop apex (solid white-contour line in Figure \ref{sji_aia_channels} and Figure \ref{dem_maps}) with respect to time. We compare the evolution of the emission at hot temperatures with that at cool temperatures, seen in  the histograms corresponding to  coronal rain pixels for these showers, not only at the apex but also along the legs. 
A steady decrease of the EM at hot temperatures ($\log T=6.3-6.5$) can be observed for both events (Shower 19$+$20 and Shower 32) up to 20~min prior (case 19+20) or right until the start of the shower (case 32), while at cool coronal temperatures ($\log T=6-6.1$) we see an increasing trend. For the case of Shower 19$+$20, the $\log T=5.8-5.9$ curves show a decrease prior to the shower event followed by a broad peak over the time the shower appears at the apex. The emission at transition region temperatures ($\log T=5.6-5.7$) is more variable, but also increases at the time of shower appearance. The AIA~304 emission, dominated by \ion{He}{II} with formation temperatures of $\approx\log T=5$, exhibits a small peak 5~min prior to the rain appearance in the SJI channels ($\log T=4-4.8$), and a larger peak that coincides with the SJI peaks. The former peak corresponds to a small rain clump in the loop that can actually be seen in all channels but is not detected by the RHT routine due to its low emission.

The EUV emission variation is interpreted as the cooling of plasma, with strong variations prior to the shower appearance probably due to continuous cooling passing through temperature ranges and also moving out of the apex. The EUV variations seen during the shower appearance are probably due to the Condensation Corona Transition Region (CCTR), as expected from numerical modelling \citep{Antolin_2022ApJ...926L..29A}. The Shower 32 behaves slightly differently than Shower 19$+$20.  The rain is seen to stay longer at the apex (40~min, almost as twice the time for shower 19+20) before falling towards the surface of the Sun. 
The cooling trend can also be clearly seen in the DEM-weighted temperature plots (Figure~\ref{em_dem_plots}, bottom). Here, as expected, the hot temperature emission dominates, so the average temperature is still coronal. 

\subsection{TNE Volume} \label{sec:TNE}
Here we aim at estimating the volume affected by TNE over the active region. Given the very good match found between the shower widths and coronal loop widths, as well as the significant loop portion occupied by showers, we use the properties of the rain clumps and showers for the TNE volume calculation. We divide this calculation into the following steps, which we perform for each channel. Since rain is dynamic and falling on average, two successive snapshots may have an overlap of rain pixels in the FOV area, depending on the clump length, velocity and instrument cadence. The length and area of this overlap is:
\begin{eqnarray}
<length\_overlap> &=& <l_{clump}> - cadence \times <v_{clump}>\label{step:1}\\
<area\_overlap> &=& <length\_overlap>\times<w_{clump}>, \label{step:2}
\end{eqnarray} 
where $<l_{clump}>$, $<w_{clump}>$ and $<v_{clump}>$ denote, respectively, the average clump's length, width and speed. 

We approximate the area in the FOV occupied by a single clump as:
\begin{equation} \label{step:3}
<area\_clump> = <l_{clump}>\times<w_{clump}>
\end{equation}
Then, the fraction of rain overlapping between 2 consecutive images is: 
\begin{equation} \label{step:4}
<fraction> = \frac{<area\_overlap>}{<area\_clump>}
\end{equation}
The number of pixels in the non-overlapping area is then given by:
\begin{equation} \label{step:5}
N_{no\_overlap} = N_{\theta_{xy}} \times (1 - <fraction>),
\end{equation}
where $N_{\theta_{xy}}$ is the number of rain pixels detected with the RHT algorithm. The number of expected shower events is given by:
\begin{equation} \label{step:6}
N_{exp\_shower} = \frac{N_{no\_overlap} \times N_{shower}}{N_{shower\_pixels}},
\end{equation}
where $N_{shower}=50$ correspond to the manually identified showers, and $N_{shower\_pixels}$ is the total number of pixels for these 50 showers.

Finally, the TNE volume is estimated as: 
\begin{equation} \label{step:7}
V_{TNE} = \pi \frac{1}{f} N_{exp\_shower} \times <l_{shower}>  \left(\frac{<w_{shower}>}{2}\right)^2,
\end{equation}
where $f$ denotes the average fraction of the loop occupied by a shower (we take $f=1/3$) and we approximate a shower as a cylinder. 

To estimate  the number of expected showers (in Equation \ref{step:6}) we used the average clump's width $<w_{clumps}>$ ($1.36\pm0.35, 0.88\pm0.27$, and $0.97\pm0.27$~Mm for the AIA~304, SJI~1400, and SJI~2796, respectively) and length $<l_{clumps}>$ ($11.53\pm7.04, 6.84\pm4.77$, and $8.99\pm6.48$~Mm for the AIA~304, SJI~1400, and SJI~2796, respectively) \citep{Sahin_rht_2022}. The number of expected shower events found are $185\pm39, 208\pm77$, and $71\pm4$ for the SJI~2796, SJI~1400, and AIA~304, respectively. The TNE volume is found to be $6.34\pm4.91 \times 10^{28}$ cm$^{3}$ for SJI~2796, $5.26\pm4.52 \times 10^{28}$ cm$^{3}$ for SJI~1400, and $2.07\pm1.71 \times 10^{28}$ cm$^{3}$ for AIA~304. 
We expect the estimated TNE volume to be a lower estimate since the coronal rain detection conditions used for the RHT routine are strict to avoid any influence from noise, while the rain intensity can sometimes be on the same order of the noise  \citep[see][for detailed information]{schad2017,Sahin_rht_2022}.

\section{Discussion and Conclusions} \label{sec:discussion}

The first major question we address is whether the coronal volume occupied by a shower can provide a proper definition or help identify what we loosely observationally attribute as a coronal loop. This question is relevant since the coronal magnetic field cannot be directly observed and the optically thin coronal emission can easily lead to misleading loop-like structure, best exemplified by the concept of `coronal veil' introduced by \citet{Malanushenko_2022ApJ...927....1M}. Furthermore, a debate exists on whether a coronal loop can be well-defined at all, since global MHD simulations suggest continuous magnetic connectivity changes and fuzzy boundaries between different magnetic field structures \citep{gudiksen2005}. On the other hand, the occurrence of a coronal rain shower points to the TNE-TI scenario \citep{antolin2020}, and therefore a specific thermodynamic evolution that is coherent over a specific bundle of magnetic field lines. Furthermore, showers can have a significant optical thickness in chromospheric and transition region lines (given the high coronal rain densities and abundance of rain clumps within a shower), thereby reducing the projection effect. 

To answer this question we conducted the first statistical investigation on the morphological and thermodynamic properties of coronal rain showers using IRIS and SDO/AIA spanning chromospheric and transition region temperatures. We found showers to be ubiquitous over an active region observed at the East limb over 5.45 hours. We manually identified 50 shower events and found their average lengths and widths on the order of 27$\pm 11.95$~Mm and 2$\pm 0.74$~Mm, respectively, with little variation across the spanned temperature range. The obtained widths are similar to the first estimates made in \citet{antolin2012} and \citet{antolin2015} for individual events. We morphologically compared the showers with the observed coronal loops that host them, seen in the EUV bands, and found good agreement regarding the widths. The obtained widths are also in agreement with previous EUV and X-ray observations of loops \citep{Aschwanden2011,Peter2013} and from 3D MHD simulations  \citep{Peter2012,Chen2014}. The significant lengths of showers (estimated to be a third of the entire loop length) and their ubiquity suggest that these structures can be used to identify coronal loop structures reliably in observations, and at least those in a TNE state. 

The morphological compatibility of our shower analysis with the studied loops in the literature also sheds light on another puzzle about loops, linked to the observed constant cross-section along their lengths. 
In agreement with EUV observations \citep{,Aschwanden2005,Fuentes2008}, little average expansion of the shower cross-section with height was found, further supporting that this effect is not apparent \citep{LopezFuentes_etal_2008ApJ...673..586L}. Although the cross-section of the flux tube may further expand over a wider volume, only a part of it (2~Mm in width) is subject to TNE, corresponding to a dense cross-section over which the temperature is homogeneous. 
This is in agreement with \citet{Peter2012} in the sense that it is specific thermodynamic cross-field distribution behind this effect. On the other hand, we observe strong expansion at low coronal heights. Between 8 and 12~Mm above the surface the width increases from 1 to 2.4~Mm, leading to an area expansion factor of 5.7, in agreement with previous individual results \citet{antolin2015}.  

We further analysed the thermodynamic evolution of a few coronal structures hosting the showers (selected based on their relative isolation in the FOV). Using DEM analysis we found global averaged cooling, in agreement with previous results \citep{Viall2012ApJ...753...35V}, but particularly evident in these coronal structures. Steady cooling and heating was found in the hot ($\log T=6.3-6.5$) and cool coronal temperatures ($\log T =6-6.1$), respectively, one hour prior to the shower events. This timescale matches the expected radiative cooling time for loops at the beginning of the TNE cycle with average temperature of $3\times10^{6}$~K and density of $10^9$~cm$^{-3}$ \citep[e.g. Eq.~18 in ][]{Antolin_Froment_2022}. Stronger variation, particularly at transition region temperatures ($\log T =5.6-5.9$) was found throughout this time, and also during the shower events. Such increased dynamics are commonly seen in active regions \citep{Ugarte-Urra2009ApJ...695..642U, Reale2014LRSP...11....4R} and match well the radiation-dominated coronal loop evolution followed by catastrophic cooling at the end of the TNE-TI cycles \citep{Antolin_2022ApJ...926L..29A}.


We have also addressed how much of the coronal volume in an active region is in a state of TNE. The importance of this question is based on the fact that TNE occurrence is linked to strongly stratified and high-frequency heating \citep{Klimchuk_2019ApJ...884...68K}, which constitute strong constraints for any heating mechanism. To answer this question we estimated the total number of showers, finding an average of 155$\pm40$ over all channels. 
Assumming that a shower occupies on average 1/3 of a coronal loop, we obtained an estimate for the average TNE volume of 4.56$\pm3.71$ $\times$10$^{28}$ cm$^{3}$. Taking the IRIS FOV as a rough estimate of the total AR volume (assuming the same length along the line-of-sight as the width of the FOV), we obtain a AR volume of 8.7 $\times$10$^{28}$ cm$^{3}$, by approximating the shape of the AR to a trapezoidal shape covering the observed coronal loops. Our TNE volume estimate is roughly half the AR volume. This is very likely a lower estimate of the TNE volume because we are not able to detect with IRIS the loops with TNE cycles without catastrophic cooling (incomplete condensations) for which the temperatures do not go down to transition region values or lower. Similarly, the probability of detecting loops with TNE cycle periods longer than the duration of our observing time sequence decreases with the period, and according to \citet{Auchere_2014AA...563A...8A} and \citet{Froment_2016PhDT.......115F}, their occurrence frequency remains roughly the same for periods between 6 and 16 hrs in active regions. Furthermore, we have used strict conditions for coronal rain detection, which ignores all rain with low emissivity. These results therefore suggest a prevalence of TNE in this AR, indicating strongly stratified and high-frequency heating on average.

\acknowledgements{\longacknowledgment}

\bibliographystyle{aasjournal}

\bibliography{sample631}{}

\begin{thebibliography}{}
\expandafter\ifx\csname natexlab\endcsname\relax\def\natexlab#1{#1}\fi
\providecommand{\url}[1]{\href{#1}{#1}}
\providecommand{\dodoi}[1]{doi:~\href{http://doi.org/#1}{\nolinkurl{#1}}}
\providecommand{\doeprint}[1]{\href{http://ascl.net/#1}{\nolinkurl{http://ascl.net/#1}}}
\providecommand{\doarXiv}[1]{\href{https://arxiv.org/abs/#1}{\nolinkurl{https://arxiv.org/abs/#1}}}

\bibitem[{{Antolin}(2020)}]{antolin2020}
{Antolin}, P. 2020, Plasma Physics and Controlled Fusion, 62, 014016,
  \dodoi{10.1088/1361-6587/ab5406}

\bibitem[{{Antolin} \& {Froment}(2022)}]{Antolin_Froment_2022}
{Antolin}, P., \& {Froment}, C. 2022, Frontiers in Astronomy and Space
  Sciences, Accepted

\bibitem[{{Antolin} {et~al.}(2022){Antolin}, {Mart{\'\i}nez-Sykora}, \&
  {{\c{S}}ahin}}]{Antolin_2022ApJ...926L..29A}
{Antolin}, P., {Mart{\'\i}nez-Sykora}, J., \& {{\c{S}}ahin}, S. 2022, \apjl,
  926, L29, \dodoi{10.3847/2041-8213/ac51dd}

\bibitem[{{Antolin} \& {Rouppe van der Voort}(2012)}]{antolin2012}
{Antolin}, P., \& {Rouppe van der Voort}, L. 2012, \apj, 745, 152,
  \dodoi{10.1088/0004-637X/745/2/152}

\bibitem[{{Antolin} {et~al.}(2015){Antolin}, {Vissers}, {Pereira}, {Rouppe van
  der Voort}, \& {Scullion}}]{antolin2015}
{Antolin}, P., {Vissers}, G., {Pereira}, T.~M.~D., {Rouppe van der Voort}, L.,
  \& {Scullion}, E. 2015, \apj, 806, 81, \dodoi{10.1088/0004-637X/806/1/81}

\bibitem[{{Aschwanden} \& {Boerner}(2011)}]{Aschwanden2011}
{Aschwanden}, M.~J., \& {Boerner}, P. 2011, \apj, 732, 81,
  \dodoi{10.1088/0004-637X/732/2/81}

\bibitem[{{Aschwanden} \& {Nightingale}(2005)}]{Aschwanden2005}
{Aschwanden}, M.~J., \& {Nightingale}, R.~W. 2005, \apj, 633, 499,
  \dodoi{10.1086/452630}

\bibitem[{{Auch{\`e}re} {et~al.}(2014){Auch{\`e}re}, {Bocchialini}, {Solomon},
  \& {Tison}}]{Auchere_2014AA...563A...8A}
{Auch{\`e}re}, F., {Bocchialini}, K., {Solomon}, J., \& {Tison}, E. 2014, \aap,
  563, A8, \dodoi{10.1051/0004-6361/201322572}

\bibitem[{{Auch{\`e}re} {et~al.}(2018){Auch{\`e}re}, {Froment}, {Soubri{\'e}},
  {Antolin}, {Oliver}, \& {Pelouze}}]{auchere2018}
{Auch{\`e}re}, F., {Froment}, C., {Soubri{\'e}}, E., {et~al.} 2018, \apj, 853,
  176, \dodoi{10.3847/1538-4357/aaa5a3}

\bibitem[{{Chen} {et~al.}(2014){Chen}, {Peter}, {Bingert}, \&
  {Cheung}}]{Chen2014}
{Chen}, F., {Peter}, H., {Bingert}, S., \& {Cheung}, M.~C.~M. 2014, \aap, 564,
  A12, \dodoi{10.1051/0004-6361/201322859}

\bibitem[{{Chen} {et~al.}(2022){Chen}, {Tian}, {Li}, {Peter}, {Chitta}, \&
  {Hou}}]{ChenH_2022AA...659A.107C}
{Chen}, H., {Tian}, H., {Li}, L., {et~al.} 2022, \aap, 659, A107,
  \dodoi{10.1051/0004-6361/202142093}

\bibitem[{{Cheung} {et~al.}(2015){Cheung}, {Boerner}, {Schrijver}, {Testa},
  {Chen}, {Peter}, \& {Malanushenko}}]{Cheung2015}
{Cheung}, M. C.~M., {Boerner}, P., {Schrijver}, C.~J., {et~al.} 2015, \apj,
  807, 143, \dodoi{10.1088/0004-637X/807/2/143}

\bibitem[{Claes \& Keppens(2021)}]{Claes:2021wg}
Claes, N., \& Keppens, R. 2021, Solar Physics, 296, 143,
  \dodoi{10.1007/s11207-021-01894-2}

\bibitem[{{De Pontieu} {et~al.}(2014){De Pontieu}, {Title}, {Lemen}, {Kushner},
  {Akin}, {Allard}, {Berger}, {Boerner}, {Cheung}, {Chou}, {Drake}, {Duncan},
  {Freeland}, {Heyman}, {Hoffman}, {Hurlburt}, {Lindgren}, {Mathur}, {Rehse},
  {Sabolish}, {Seguin}, {Schrijver}, {Tarbell}, {W{\"u}lser}, {Wolfson},
  {Yanari}, {Mudge}, {Nguyen-Phuc}, {Timmons}, {van Bezooijen}, {Weingrod},
  {Brookner}, {Butcher}, {Dougherty}, {Eder}, {Knagenhjelm}, {Larsen},
  {Mansir}, {Phan}, {Boyle}, {Cheimets}, {DeLuca}, {Golub}, {Gates}, {Hertz},
  {McKillop}, {Park}, {Perry}, {Podgorski}, {Reeves}, {Saar}, {Testa}, {Tian},
  {Weber}, {Dunn}, {Eccles}, {Jaeggli}, {Kankelborg}, {Mashburn}, {Pust},
  {Springer}, {Carvalho}, {Kleint}, {Marmie}, {Mazmanian}, {Pereira}, {Sawyer},
  {Strong}, {Worden}, {Carlsson}, {Hansteen}, {Leenaarts}, {Wiesmann},
  {Aloise}, {Chu}, {Bush}, {Scherrer}, {Brekke}, {Martinez-Sykora}, {Lites},
  {McIntosh}, {Uitenbroek}, {Okamoto}, {Gummin}, {Auker}, {Jerram}, {Pool}, \&
  {Waltham}}]{depontieu2014}
{De Pontieu}, B., {Title}, A.~M., {Lemen}, J.~R., {et~al.} 2014, \solphys, 289,
  2733, \dodoi{10.1007/s11207-014-0485-y}

\bibitem[{{DeForest}(2007)}]{DeForest2007}
{DeForest}, C.~E. 2007, \apj, 661, 532, \dodoi{10.1086/515561}

\bibitem[{Dudok~de Wit {et~al.}(2013)Dudok~de Wit, Moussaoui, Guennou,
  Auch{\`e}re, Cessateur, Kretzschmar, Vieira, \&
  Goryaev}]{Dudok-de-Wit:2013uw}
Dudok~de Wit, T., Moussaoui, S., Guennou, C., {et~al.} 2013, Solar Physics,
  283, 31, \dodoi{10.1007/s11207-012-0142-2}

\bibitem[{{Fang} {et~al.}(2013){Fang}, {Xia}, \& {Keppens}}]{fang2013}
{Fang}, X., {Xia}, C., \& {Keppens}, R. 2013, \apjl, 771, L29,
  \dodoi{10.1088/2041-8205/771/2/L29}

\bibitem[{{Fang} {et~al.}(2015){Fang}, {Xia}, {Keppens}, \& {Van
  Doorsselaere}}]{Fang_2015ApJ...807..142F}
{Fang}, X., {Xia}, C., {Keppens}, R., \& {Van Doorsselaere}, T. 2015, \apj,
  807, 142, \dodoi{10.1088/0004-637X/807/2/142}

\bibitem[{{Froment}(2016)}]{Froment_2016PhDT.......115F}
{Froment}, C. 2016, PhD thesis, Institut d'Astrophysique Spatiale

\bibitem[{{Froment} {et~al.}(2020){Froment}, {Antolin}, {Henriques},
  {Kohutova}, \& {Rouppe van der Voort}}]{froment2020}
{Froment}, C., {Antolin}, P., {Henriques}, V.~M.~J., {Kohutova}, P., \& {Rouppe
  van der Voort}, L.~H.~M. 2020, \aap, 633, A11,
  \dodoi{10.1051/0004-6361/201936717}

\bibitem[{{Froment} {et~al.}(2015){Froment}, {Auch{\`e}re}, {Bocchialini},
  {Buchlin}, {Guennou}, \& {Solomon}}]{Froment_2015ApJ...807..158F}
{Froment}, C., {Auch{\`e}re}, F., {Bocchialini}, K., {et~al.} 2015, \apj, 807,
  158, \dodoi{10.1088/0004-637X/807/2/158}

\bibitem[{{Froment} {et~al.}(2018){Froment}, {Auch{\`e}re}, {Miki{\'c}},
  {Aulanier}, {Bocchialini}, {Buchlin}, {Solomon}, \&
  {Soubri{\'e}}}]{froment2018ApJ}
{Froment}, C., {Auch{\`e}re}, F., {Miki{\'c}}, Z., {et~al.} 2018, \apj, 855,
  52, \dodoi{10.3847/1538-4357/aaaf1d}

\bibitem[{{Gudiksen} \& {Nordlund}(2005)}]{gudiksen2005}
{Gudiksen}, B.~V., \& {Nordlund}, {\r{A}}. 2005, \apj, 618, 1031,
  \dodoi{10.1086/426064}

\bibitem[{{Jing} {et~al.}(2016){Jing}, {Xu}, {Cao}, {Liu}, {Gary}, \&
  {Wang}}]{Jing_2016NatSR...624319J}
{Jing}, J., {Xu}, Y., {Cao}, W., {et~al.} 2016, Scientific Reports, 6, 24319,
  \dodoi{10.1038/srep24319}

\bibitem[{{Klimchuk}(2000)}]{Klimchuk2000SoPh..193...53K}
{Klimchuk}, J.~A. 2000, \solphys, 193, 53, \dodoi{10.1023/A:1005210127703}

\bibitem[{{Klimchuk} \& {Luna}(2019)}]{Klimchuk_2019ApJ...884...68K}
{Klimchuk}, J.~A., \& {Luna}, M. 2019, \apj, 884, 68,
  \dodoi{10.3847/1538-4357/ab41f4}

\bibitem[{{Lemen} {et~al.}(2012){Lemen}, {Title}, {Akin}, {Boerner}, {Chou},
  {Drake}, {Duncan}, {Edwards}, {Friedlaender}, {Heyman}, {Hurlburt}, {Katz},
  {Kushner}, {Levay}, {Lindgren}, {Mathur}, {McFeaters}, {Mitchell}, {Rehse},
  {Schrijver}, {Springer}, {Stern}, {Tarbell}, {Wuelser}, {Wolfson}, {Yanari},
  {Bookbinder}, {Cheimets}, {Caldwell}, {Deluca}, {Gates}, {Golub}, {Park},
  {Podgorski}, {Bush}, {Scherrer}, {Gummin}, {Smith}, {Auker}, {Jerram},
  {Pool}, {Soufli}, {Windt}, {Beardsley}, {Clapp}, {Lang}, \&
  {Waltham}}]{lemen2012}
{Lemen}, J.~R., {Title}, A.~M., {Akin}, D.~J., {et~al.} 2012, \solphys, 275,
  17, \dodoi{10.1007/s11207-011-9776-8}

\bibitem[{{Li} {et~al.}(2018){Li}, {Zhang}, {Peter}, {Chitta}, {Su}, {Xia},
  {Song}, \& {Hou}}]{LiL_2018ApJ...864L...4L}
{Li}, L., {Zhang}, J., {Peter}, H., {et~al.} 2018, \apjl, 864, L4,
  \dodoi{10.3847/2041-8213/aad90a}

\bibitem[{{Li} {et~al.}(2022){Li}, {Keppens}, \&
  {Zhou}}]{LiX_2022ApJ...926..216L}
{Li}, X., {Keppens}, R., \& {Zhou}, Y. 2022, \apj, 926, 216,
  \dodoi{10.3847/1538-4357/ac41cd}

\bibitem[{{L{\'o}pez Fuentes} {et~al.}(2008{\natexlab{a}}){L{\'o}pez Fuentes},
  {D{\'e}moulin}, \& {Klimchuk}}]{Fuentes2008}
{L{\'o}pez Fuentes}, M.~C., {D{\'e}moulin}, P., \& {Klimchuk}, J.~A.
  2008{\natexlab{a}}, \apj, 673, 586, \dodoi{10.1086/523928}

\bibitem[{{L{\'o}pez Fuentes} {et~al.}(2008{\natexlab{b}}){L{\'o}pez Fuentes},
  {D{\'e}moulin}, \& {Klimchuk}}]{LopezFuentes_etal_2008ApJ...673..586L}
---. 2008{\natexlab{b}}, \apj, 673, 586, \dodoi{10.1086/523928}

\bibitem[{{Malanushenko} {et~al.}(2022){Malanushenko}, {Cheung}, {DeForest},
  {Klimchuk}, \& {Rempel}}]{Malanushenko_2022ApJ...927....1M}
{Malanushenko}, A., {Cheung}, M.~C.~M., {DeForest}, C.~E., {Klimchuk}, J.~A.,
  \& {Rempel}, M. 2022, \apj, 927, 1, \dodoi{10.3847/1538-4357/ac3df9}

\bibitem[{{Morgan} \& {Druckm{\"u}ller}(2014)}]{Morgan2014}
{Morgan}, H., \& {Druckm{\"u}ller}, M. 2014, \solphys, 289, 2945,
  \dodoi{10.1007/s11207-014-0523-9}

\bibitem[{{M{\"u}ller} {et~al.}(2003){M{\"u}ller}, {Hansteen}, \&
  {Peter}}]{Muller_2003AA...411..605M}
{M{\"u}ller}, D.~A.~N., {Hansteen}, V.~H., \& {Peter}, H. 2003, \aap, 411, 605,
  \dodoi{10.1051/0004-6361:20031328}

\bibitem[{{Oliver} {et~al.}(2014){Oliver}, {Soler}, {Terradas}, {Zaqarashvili},
  \& {Khodachenko}}]{Oliver_2014ApJ...784...21O}
{Oliver}, R., {Soler}, R., {Terradas}, J., {Zaqarashvili}, T.~V., \&
  {Khodachenko}, M.~L. 2014, \apj, 784, 21, \dodoi{10.1088/0004-637X/784/1/21}

\bibitem[{{Pesnell} {et~al.}(2012){Pesnell}, {Thompson}, \&
  {Chamberlin}}]{pesnell2012}
{Pesnell}, W.~D., {Thompson}, B.~J., \& {Chamberlin}, P.~C. 2012, \solphys,
  275, 3, \dodoi{10.1007/s11207-011-9841-3}

\bibitem[{{Peter} \& {Bingert}(2012)}]{Peter2012}
{Peter}, H., \& {Bingert}, S. 2012, \aap, 548, A1,
  \dodoi{10.1051/0004-6361/201219473}

\bibitem[{{Peter} {et~al.}(2013){Peter}, {Bingert}, {Klimchuk}, {de Forest},
  {Cirtain}, {Golub}, {Winebarger}, {Kobayashi}, \& {Korreck}}]{Peter2013}
{Peter}, H., {Bingert}, S., {Klimchuk}, J.~A., {et~al.} 2013, \aap, 556, A104,
  \dodoi{10.1051/0004-6361/201321826}

\bibitem[{{Reale}(2014)}]{Reale2014LRSP...11....4R}
{Reale}, F. 2014, Living Reviews in Solar Physics, 11, 4,
  \dodoi{10.12942/lrsp-2014-4}

\bibitem[{{Sahin} {et~al.}(in preparation){Sahin}, {Antolin}, {Froment}, \&
  {Schad}}]{Sahin_rht_2022}
{Sahin}, S., {Antolin}, P., {Froment}, C., \& {Schad}, T. in preparation, \apj

\bibitem[{{Schad}(2017)}]{schad2017}
{Schad}, T. 2017, \solphys, 292, 132, \dodoi{10.1007/s11207-017-1153-9}

\bibitem[{{Schrijver}(2001)}]{schrijver2001}
{Schrijver}, C.~J. 2001, \solphys, 198, 325, \dodoi{10.1023/A:1005211925515}

\bibitem[{{Scullion} {et~al.}(2016){Scullion}, {Rouppe van der Voort},
  {Antolin}, {Wedemeyer}, {Vissers}, {Kontar}, \& {Gallagher}}]{scullion2016}
{Scullion}, E., {Rouppe van der Voort}, L., {Antolin}, P., {et~al.} 2016, \apj,
  833, 184, \dodoi{10.3847/1538-4357/833/2/184}

\bibitem[{{Ugarte-Urra} {et~al.}(2009){Ugarte-Urra}, {Warren}, \&
  {Brooks}}]{Ugarte-Urra2009ApJ...695..642U}
{Ugarte-Urra}, I., {Warren}, H.~P., \& {Brooks}, D.~H. 2009, \apj, 695, 642,
  \dodoi{10.1088/0004-637X/695/1/642}

\bibitem[{{Viall} \& {Klimchuk}(2012)}]{Viall2012ApJ...753...35V}
{Viall}, N.~M., \& {Klimchuk}, J.~A. 2012, \apj, 753, 35,
  \dodoi{10.1088/0004-637X/753/1/35}

\bibitem[{{Watko} \& {Klimchuk}(2000)}]{Watko2000}
{Watko}, J.~A., \& {Klimchuk}, J.~A. 2000, \solphys, 193, 77,
  \dodoi{10.1023/A:1005209528612}

\end{thebibliography}



\end{document}